\documentclass{llncs}
\usepackage{proof}
\usepackage{amssymb}
\usepackage{amsmath}
\usepackage{stmaryrd}
\usepackage[all]{xy}
\RequirePackage{txfonts}
\DeclareMathAlphabet{\mathsl}{OT1}{cmr}{m}{sl}

\usepackage{color}
\usepackage{url}
\usepackage{ifpdf}
\usepackage{lineno}
\usepackage{balance}
\usepackage{graphicx}
\usepackage{subcaption}
\usepackage{lipsum}
\usepackage{wrapfig}
\usepackage{xspace}
\usepackage{times}
\usepackage{multirow}
\usepackage{booktabs}

\ifpdf
\usepackage[colorlinks=true,linkcolor=blue,citecolor=blue]{hyperref}
\fi
\usepackage{xypic}

\usepackage{cite}
\usepackage[multiple]{footmisc}

\newcommand{\red}[1]{\textcolor{red}{#1} }
\long\def\comment#1{\relax}
\newcommand{\eg}{{\em e.g.}}
\newcommand{\ie}{{\em i.e.}}
\newcommand{\etal}{\emph{et al.}}

\newcommand{\Bscr}{\mathcal{B}}
\newcommand{\Pscr}{\mathcal{P}}

\newcommand{\Iscr}{\mathcal{I}}

\newcommand{\Cscr}{\mathcal{C}}

\newcommand{\Nscr}{\mathcal{N}}

\newcommand{\Tscr}{\mathcal{T}}

\newcommand{\Gscr}{\mathcal{G}}
\newcommand{\TC}{\mathcal{TC}}
\newcommand{\PL}{\mathcal{PL}}
\newcommand{\GPL}{\mathcal{GPL}}

\newcommand\sk{\ensuremath{\mathsf{sk}}\xspace}
\newcommand\pk{\ensuremath{\mathsf{pk}}\xspace}
\newcommand\enc{\ensuremath{\mathsf{enc}}\xspace}
\newcommand\n{\ensuremath{\mathsf{n}}\xspace}
\renewcommand\k{\ensuremath{\mathsf{k}}\xspace}
\newcommand\ti{\ensuremath{\mathsf{ti}}\xspace}
\renewcommand\b{\ensuremath{\mathsf{b}}\xspace}
\renewcommand\c{\ensuremath{\mathsf{c}}\xspace}
\renewcommand\t{\ensuremath{\mathsf{t}}\xspace}
\newcommand\pl{\ensuremath{\mathsf{pl}}\xspace}
\renewcommand\r{\ensuremath{\mathsf{r}}\xspace}
\newcommand\m{\ensuremath{\mathsf{m}}\xspace}
\renewcommand\v{\ensuremath{\mathsf{v}}\xspace}
\newcommand\p{\ensuremath{\mathsf{p}}\xspace}

\newcommand\tv{\ensuremath{\mathsf{tt}}\xspace}
\newcommand\lst[1]{\ensuremath{\{#1\}}\xspace}
\newcommand\cur{\ensuremath{\mathsf{cur}}\xspace}
\newcommand\floor{\ensuremath{\mathsf{floor}}\xspace}
\newcommand\ceiling{\ensuremath{\mathsf{ceiling}}\xspace}
\newcommand\tr{\ensuremath{\mathsf{tr}}\xspace}
\newcommand\tc{\ensuremath{\mathsf{tc}}\xspace}
\newcommand\new{\ensuremath{\mathsf{new}}\xspace}
\newcommand\myif[3]{\ensuremath{\mathsf{if}~#1~\mathsf{then}~#2~\mathsf{else}~#3}\xspace}
\newcommand\td[2]{\ensuremath{\mathsf{td}(#1,#2)}\xspace}
\newcommand\key[1]{\ensuremath{\mathsf{#1}}\xspace}

\newcommand{\tup}[1]{\langle#1\rangle}

\renewenvironment{paragraph}[1]{\smallskip\noindent{\it #1}~}

\usepackage{tweaklist}
\renewcommand{\enumhook}{
	\setlength{\topsep}{0em}
	\setlength{\itemsep}{0em}
	\setlength{\partopsep}{0em}
	\setlength{\parskip}{0em}
	\setlength{\parsep}{0em}
	\addtolength{\leftmargin}{-0.75em}
}
\renewcommand{\itemhook}{
	\setlength{\topsep}{0em}
	\setlength{\itemsep}{0em}
	\setlength{\partopsep}{0em}
	\setlength{\parskip}{0em}
	\setlength{\parsep}{0em}
	\addtolength{\leftmargin}{-0.75em}
}

\newcommand\lra{\longrightarrow}

\title{Towards the Automated Verification of Cyber-Physical Security Protocols: Bounding the Number of Timed Intruders}

\author{Vivek Nigam\inst{1}, Carolyn Talcott\inst{2} and Abra\~ao Aires Urquiza\inst{1}}
\institute{
Federal University of Para\'iba, Brazil,
\email{vivek@ci.ufpb.br, abraauc@gmail.com}
\and
SRI International, USA, \email{clt@csl.sri.com} 
}

\begin{document}
\maketitle


\begin{abstract}
Timed Intruder Models have been proposed for the verification of Cyber-Physical Security Protocols (CPSP) amending the traditional Dolev-Yao intruder to obey the physical restrictions of the environment. Since to learn a message, a Timed Intruder  needs to wait for a message to arrive, mounting an attack may depend on where Timed Intruders are. It may well be the case that in the presence of a great number of intruders there is no attack, but there is an attack in the presence of a small number of well placed intruders. Therefore, a major challenge for the automated verification of CPSP is to determine how many Timed Intruders to use and where should they be placed. This paper answers this question by showing it is enough to use the same number of Timed Intruders as the number of participants. We also report on some preliminary experimental results in discovering attacks in CPSP.

\end{abstract}


\section{Introduction}
\label{sec:intro}
The Dolev-Yao intruder model is one of the cornerstones for the success of protocol verification being used in most verification tools. The protocol security literature contains a number of properties about the Dolev-Yao intruder, many of them vital for automated verification. For instance, it has been shown that protocol security verification is complete when considering only a single Dovel-Yao intruder in the following sense: if there is an attack in the presence of one or more (colluding) Dolev-Yao intruders, then the same attack with a single Dolev-Yao intruder is possible. Such result greatly simplifies the implementation of tools as it is enough to use only one Dolev-Yao intruder.

However, for the important class of Cyber-Physical Security Protocols (CPSP), the Dolev-Yao intruder model is not suitable. CSPS normally rely on the physical properties of the environment where sessions are carried out to establish some physical properties. For example, Distance Bounding Protocols are used to infer an upper-bound on the distance between two players $V$, the verifier, and $P$, the prover. It works as follows:
\[
  \begin{array}{ll}
    V \lra P : m\\
    P \lra V : m'
  \end{array}
\]
The verifier sends a challenge $m$ remembering the time $t_1$, when this message is sent. The prover responds to the challenge, $m'$, and by measuring the round-trip time of the challenge response round, the verifier can compute (using assumptions on the transmission channel used) an upper bound on the distance to the prover.

It is easy to check that the Dolev-Yao intruder is not suitable for CPSP verification, as the Dolev-Yao intruder does not obey the physical properties of the system. As the Dolev-Yao intruder controls the network, he can receive the challenge $m$ and instantaneously respond $m'$ to the verifier's challenge. There have been, therefore, proposals to amend the Dolev-Yao intruder model to CSPS~\cite{kanovich14fccfcs,basin11iss} in the form of Timed Dolev-Yao models. These have been used to prove general decidability of important properties of CSPS~\cite{kanovich15post,basin11iss} and prove the security of protocols using theorem provers. 

\begin{figure}[t]
\begin{center}
\includegraphics[width=0.75\textwidth]{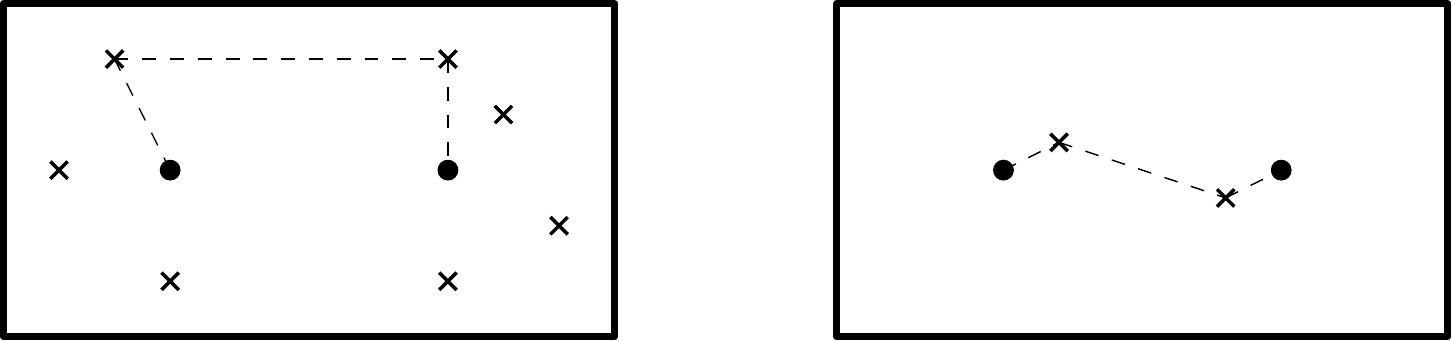} 
\end{center}
\vspace{-3mm}
\caption{The dots are protocol participants and the crosses are intruders.}
\vspace{-5mm}
\label{fig:place-intruders}
\end{figure}

In contrast with the traditional Dolev-Yao intruder, who is the whole network, a timed intruder is placed at some location and in order to learn a message,  must wait until the message arrives to that location. A consequence of this is that a greater number of colluding intruders may not do as much damage as a smaller number of intruders that are better placed. For example, consider Figure~\ref{fig:place-intruders}. With a distribution of intruders shown to the left, there may not be an attack as it might take too long for intercepting and forwarding messages among intruders (illustrated by the dashed lines), while there may be an attack with the distribution of intruders shown to the right.

The main contribution of this paper is to answer the question: \emph{How many intruders are enough for verification and where should they be placed?} We prove that it is enough to consider one intruder per protocol participant,  thus bounding the number of timed intruders. This result greatly simplifies automated CSPS verification as the specifier no longer has to guess how many timed intruders to consider and where to place them.

Our second contribution is a general specification language, which extends strand spaces~\cite{thayer99jcs} by allowing for the symbolic representation of time. Instead of instantiating time variables and time constraints with explicit values, the semantics of our language accumulates symbolic time constraints. An execution using symbolic time constraints corresponds to a set of possible concrete executions, considerably reducing state-space. We implemented a prototype of our language in Maude~\cite{clavel-etal-07maudebook} with SMT support. Our preliminary experiments show that it is possible to find attacks traversing few states. While we do not claim (yet) to have a complete tool, our first results are promising.

This paper is structured as follows: Section~\ref{sec:language} specifies the syntax of our protocol specification language and its semantics extending Strand Spaces~\cite{thayer99jcs}. We introduce the Timed Intruder Model in Section~\ref{sec:intruder}. Section~\ref{sec:completeness} contains the definition of the Timed Intruder Completeness problem and a solution to it. We revisit some examples in Section~\ref{sec:examples} briefly commenting on our prototype implementation. Finally we conclude by reviewing related and future work in Section~\ref{sec:related}.






\vspace{-2mm}
\section{A Specification Language for Cyber-Physical Security Protocols}
\label{sec:language}
\vspace{-2mm}

We start first by specifying the syntax of our CPSP specification language with symbolic time variables and symbolic time constraints. We exemplify the specification of protocols using our language. Then, we formalize the operational semantics of our language by extending Strand Spaces~\cite{thayer99jcs} to include time variables.

\vspace{-2mm}
\subsection{Syntax}
\vspace{-2mm}
\paragraph{Message Expressions}
We assume a message signature $\Sigma$ of constants, and function symbols. 
Constants include nonces, symmetric keys and player names.
The set of messages is constructed as usual using constants,  variables 
and at least the following function symbols:
\[
\begin{array}{l@{\qquad}l}
  \sk(\p) & \textrm{Denoting the secret key of the player $\p$;}\\
  \pk(\p) & \textrm{Denoting the public key of the player $\p$;}\\
  \enc(\m,\k) & \textrm{Encryption function denoting the encryption of $m$ using key $k$;}\\
  \lst{\m_1,\m_2,\ldots,\m_n} & \textrm{Tuple function denoting a list of messages $\m_1,\m_2,\ldots,\m_n$;}\\
\end{array}  
\]
where $\c_1, \c_2, \ldots$ range over constants, 
$\n_1,\ldots,\n_n$ range over nonces,
$\k_1,\k_2,\ldots$ range over symmetric keys, 
$\p_1, \p_2 \ldots$ range over player names,
$\v_1,\v_2, \ldots$ range over variables, and
$\m_1,\m_2,\ldots$ range over messages.
For example, the message $\enc(\lst{\v_1, \enc(\c,\k)},\pk(\p))$ denotes the encryption using the public key of \p of the pair of messages $\v_1$ (not instantiated) and $\enc(\c,\k)$. We define $ (\pk(\p))^{-1} = \sk(\p)$ and $\k^{-1} = k$ if $\k$ is a symmetric key. We also write interchangeably the singleton tuple $\{\m\}$ and $\m$.

For a given scenario with some protocol session instances, we are going to distinguish the players that are participating in the protocol sessions, \eg, as verifiers and as provers, which we call \emph{protocol participants} 
(briefly \emph{participants}), from the \emph{Timed Intruders} which are not participating explicitly in the protocol sessions in the given scenario, but are simply manipulating messages and possibly interacting with the participants. 
The symbols $\p_1, \p_2 \ldots$ will range over participant names
while  $\ti_1, \ti_2, \ldots$ will range over the names of such Timed Intruders. 

\paragraph{Time Expressions}
We also assume a time signature $\Xi$ which is disjoint to the message alphabet $\Sigma$. It contains:
\[
\begin{array}{l@{\quad}l}
  \r_1,\r_2,\ldots & \textrm{A set of numbers;}\\
  \tv_1,\tv_2,\ldots,& \textrm{A set of time variables including the special variable $\cur$};\\
  +,-,\times,/, \floor,\ceiling, \ldots & \textrm{A set of arithmetic symbols and other function symbols.}
\end{array}  
\]
\emph{Time Expressions} are constructed inductively by applying arithmetic symbols to time expressions. For example $\ceiling((2 + \tv + \cur )/ 10)$ is a Time Expression. The symbols $\tr_1,\tr_2, \ldots$ range over Time Expressions.
We do not constrain the set of numbers and function symbols in $\Xi$. However, in practice, we allow only the symbols supported by the SMT solver used. All examples in this paper will contain SMT supported symbols (or equivalent). Finally, the time variable $\cur$ will be a keyword in our protocol specification language denoting the current global time.

\begin{definition}[Symbolic Time Constraints] Let $\Xi$ be a time signature.
The set of symbolic time constraints is constructed using time expressions. 
Let  
 $\tr_1,\tr_2$ be time expressions, then 
\[
\begin{array}{l}
  \tr_1 = \tr_2, \qquad \tr_1 \geq \tr_2 \qquad \tr_1 > \tr_2, \qquad \tr_1 < \tr_2, \quad \textrm{and} \quad \tr_1 \leq \tr_2
\end{array}
\]
are Symbolic Time Constraints.
\end{definition}
For example, $\cur + 10 < \floor(\tv - 5)$ is a Time Constraint. 
The symbols $\tc_1, \tc_2, \ldots$ will range over Time Constraints.

Finally, we let $\b_1, \b_2, \ldots$, range over 
boolean expressions,  which include timed comparison constraints. 
We also allow for checking whether two messages $\m_1$ and $\m_2$ can be unified, \eg, $\lst{\v_1,\v_2} :=: \lst{\p_1, \k_1}$ evaluates to true as they can be unified by the substitution $\{\v_1 \mapsto \p_1,\v_2 \mapsto \k_1\}$. 

\begin{definition} [Timed Protocols]
The set of Timed Protocols, $\PL$, is composed of Timed Protocol Roles, $\pl$, which are constructed by using commands as specified by the following grammar, where $\b$ is a boolean expression:
\[
  \begin{array}{l@{\quad}c@{\quad}l@{\qquad}l}
    \pl & := & \mathsf{nil} & \textrm{Empty Protocol}\\
    && \mid (\new~ \v), \mathsf{pl} & \textrm{Fresh Constant}\\
    && \mid (+ \m), \mathsf{pl} & \textrm{Message Output}\\
    && \mid (+ \m ~\#~ \tc), \mathsf{pl} &  \textrm{Timed Message Output}\\
    && \mid (- \m), \mathsf{pl} & \textrm{Message Input}\\
    && \mid (- \m ~\#~ \tc), \mathsf{pl} & \textrm{Timed Message Input}\\
    && \mid (\myif{\b}{\mathsf{pl}_1}{\mathsf{pl}_2}) & \textrm{Conditional}\\
    && \mid (\myif{\b~\#~\tc}{\mathsf{pl}_1}{\mathsf{pl}_2}) & \textrm{Timed Conditional}
  \end{array}
\]
\end{definition}

We explain some examples intuitively before we formalize the semantics of our language in the following section. We will elide $\mathsf{nil}$ whenever it is clear from the context.

\begin{example}
\label{ex:distance-bounding}
  The following program specifies the verifier of a (very simple) distance bounding protocol:
  \[
  \begin{array}{l}
   (\new~\v),(+ \v~\#~\tv = \cur), (- \v~\#~\cur \leq \tv + 4) 
  \end{array}    
  \]
  It creates a fresh constant and sends it to the prover, 
	remembering the current global time by assigning it to the time variable $\tv$. Finally, when it receives the response $\v$ it checks whether the current time is less than $\tv + 4$.

\end{example}


\begin{example}
\label{ex:passport}
Timed conditionals can be used to specify the duration of operations, such as checking whether some message is of a given form. In practice, the duration of these operations can be measured empirically to obtain a finer analysis of the protocol~\cite{chothia15fcds}.

For example, consider the following protocol role: 
\[
\begin{array}{l}
  (\new~\v),(+ \v),(- \{\v_{enc},\v_{mac}\}~\#~\tv_0 = \cur),\\
  \key{if}~ (\v_{mac} :=: \enc(\v_{enc},\k_M))~ \# ~ \tv_1 = \tv_0 + \tv_{Mac}\\
  \key{then}~ (\key{if}~(\v_{enc} :=: \enc(\v,\k_E)) ~\# ~ \tv_2 = \tv_1 + \tv_{Enc})\\
  \qquad ~\key{then} ~ (+ done ~\#~ \cur = \tv_2)~\key{else}~ (+ error ~\#~ \cur = \tv_2))\\
  \key{else}~(+ error ~\#~ \cur = \tv_1)
\end{array}  
\]
This role creates a fresh value $\v$ and sends it. Then it is expecting a pair of two messages $\v_{mac}$ and $\v_{enc}$, remembering at time variable $\tv_0$ when this message is received. It then checks whether the first component $\v_{mac}$ is of the form $\enc(\v_{enc},\k_M))$, \ie, it is the correct MAC. This operation  takes $\tv_{mac}$ time units. The time variable $\tv_1$ is equal to the time $\tv_0 + \tv_{mac}$, \ie, the time when the message was received plus the MAC check duration. If the MAC is not correct, an $error$ message is sent exactly at time $\tv_1$. Otherwise, if the first component, $\v_{MAC}$, is as expected, the role checks whether the second component, $\v_{enc}$, is an encryption of the form $\enc(\v,\k_E))$, which takes (a longer) time $\tv_{enc}$. If so it sends the $done$ message, otherwise the $error$ message, both at time $\tv_2$ which is $\tv_1 + \tv_{enc}$.
\end{example}

We will need to identify a particular command in a Timed Protocol Role. We use a string of the form $i_1.i_2.i_3.\ldots.i_n$, called position and denoted by $\bar{i}$, where each $i_j \in \{1,2\}$ to specify a path in the control flow of the Timed Protocol. For example, $1.1.1.1.2$ in Example~\ref{ex:passport} leads to $(+ error ~\#~ \cur = \tv_1)$. We denote by $\mathcal{PS}(\pl)$ the set of strings representing the paths in the Timed Protocol Role $\pl$.

\subsection{Timed Strand Spaces and Bundles}

We formalize the semantics of Timed Protocols by extending Strand Spaces and Bundles~\cite{thayer99jcs} to include time constraints and a network topology. 


\paragraph{Network Topology} Messages take time to travel between agents, both honest players and intruders. The network model is specified by representing the time a message needs to travel from any agent $a$ to any agent $b$, specified by $\td{a}{b}$ using a function that takes two names and returns a number.\footnote{Here we are assuming that two agents share a single transmission channel. We leave to future work how to incorporate different transmission channels. One way to do so is to add another parameter to $\key{td}$, which would imply the addition of more axioms.} Typically, $\td{a}{a} = 0$, that is the time for a message sent from a player to reach himself is 0, but we do not need to enforce this. We also assume the following axiom for all players $a,a_1, \ldots, a_n,a'$ (with $1 \leq n$):
\begin{equation}
  \td{a}{a'} \leq \td{a}{a_1} + \td{a_1}{a_2} + \cdots + \td{a_n}{a'}
\label{eq:geometry}
\end{equation}
That is, it is faster for a message to travel directly from $a$ to $a'$, then to first travel through $a_1,\ldots,a_n$. This is similar to the usual triangle inequality in basic geometry.

A given scenario with some protocol session instances includes the protocol participants (or simply participants), $\Pscr = \{\p_1, \ldots, \p_n\}$ and a set of Timed Intruders $\Iscr = \{\ti_1, \ldots,\ti_m\}$, who may be manipulating messages. The Network Topology is composed by two disjoint functions $\key{td} = \key{td}_\Pscr \uplus \key{td}_\Iscr$ defined as follows:
\[
\td{a}{b} = 
  \left\{ \begin{array}{ll}
    \key{td}_\Pscr(a,b) & \textrm{if $a,b \in \Pscr$}\\
    \key{td}_\Iscr(a,b) & \textrm{otherwise}
  \end{array}\right.
\]
Thus, $\key{td}_\Pscr$ specifies the time messages take to travel among participants, while $\key{td}_\Iscr$ specifies the time messages take to travel between Timed Intruders, between a Timed Intruder and a participant and between a participant and a Timed Intruder. 

The following definitions extend Strands and Bundles to include time variables capturing the semantics of Timed Protocols.


\begin{definition}
A Timed Strand Space is set $\Pi$ and a trace mapping $tr : \Pi \lra \Pscr \times \GPL$, where $\Pscr$ is the set of player names $\{\p_1, \ldots, \p_n\}$ and $\GPL$ is the set of Ground Timed Protocol Roles. We denote by $tr(s)_1$ the player name and $tr(s)_2$ the Timed Protocol Role of a strand $s \in \Pi$.
\end{definition}

For the remainder we fix a Timed Strand Space  $[\Pi,tr]$.

\begin{definition} The Timed Strand Space Graph, $\Gscr = \tup{\Nscr,\Rightarrow \cup \rightarrow}$, has nodes
$\Nscr$ and edges $\Rightarrow$ and $\rightarrow$ as defined below.
\begin{enumerate}
  \item A node $n$ is a tuple $\tup{\p,s,\bar{i}}@\tv$ with $s \in \Pi$, $\p = tr(s)_1$, $\bar{i}\in \mathcal{PS}(tr(s)_2)$ is a string identifying a command in the Timed Protocol, and $\tv$ is a time variable timestamping the node $n$. The set of nodes is denoted by $\Nscr$;
  \item If $n = \tup{\p,s,\bar{i}}@\tv$, we denote by $term(n)$, the command at position $\bar{i}$ in $tr(s)_2$;
  \item If $n_1 = \tup{\p,s,\bar{i}}@\tv_1$ and $n_2 = \tup{\p,s,\bar{i}.j}@\tv_2$ are in $\Nscr$, then there is an edge $n_1 \Rightarrow n_2$;
  \item For two nodes $\n_1, \n_2 \in \Nscr$, there is an edge $n_1 \rightarrow n_2$ if $term(n_1)$ is of the form $+\m$ or $+\m~\#~\tc_1$ and $term(n_2)$ is of the form $-\m$ or $-\m~\#~\tc_2$;
  \item If a node $n \in \Nscr$,  $term(n) = \new~\c$, then $\c$ originates on $n$, that is, all nodes $n'$ such that $term(n')$ contains $\c$ are such that $n~(\Rightarrow \cup \rightarrow)^* ~n'$, where $(\cdot)^*$ is the reflexive and transitive closure operator.
\end{enumerate}
\end{definition}

\begin{definition}
\label{def:timed-constraint-set} 
Let $\key{td}$ be a Network Topology and let $\Cscr = \tup{\Nscr_\Cscr,\rightarrow_\Cscr \cup \Rightarrow_\Cscr}$ be a subgraph of $\Gscr = \tup{\Nscr,\Rightarrow \cup \rightarrow}$. 
The Timed Constraint Set of $\Cscr$ over $\key{td}$, denoted by $\TC(\Cscr,\key{td})$, is the smallest set of Time Constraints specified as follows:
\begin{enumerate}
  \item If $n = \tup{\p,s,\bar{i}}@\tv \in \Nscr_\Cscr$, such that $term(n)$ is of the form $\pm \m~\#~\tc$ or $\key{myif}~{\b~\#~\tc}$, then $\tc' \in \TC(\Cscr,\key{td})$ where $\tc'$ is the Time Constraint obtained by replacing $\cur$ by $\tv$;
  \item If $\tup{\p,s,\bar{i}}@\tv_1 \Rightarrow_\Cscr \tup{\p,s,\bar{i}.j}@\tv_2$, then $\tv_2 \geq \tv_1 \in \TC(\Cscr,\key{td})$;
  \item If $\tup{\p_1,s_1,\bar{i}_1}@\tv_1 \rightarrow_\Cscr \tup{\p_2,s_2,\bar{i}_2}@\tv_2$, then $\tv_2 \geq \tv_1 + \td{\p_1}{\p_2} \in \TC(\Cscr,\key{td})$.
\end{enumerate}
\end{definition}

A Timed Bundle is a subset of the Timed Strand space graph.

\begin{definition} Let $\key{td}$ be a Network Topology. Let $\rightarrow_\Cscr \subseteq \rightarrow$ and $\Rightarrow_\Cscr \subseteq \Rightarrow$ and suppose $\Cscr = \tup{\Nscr_\Cscr,\rightarrow_\Cscr \cup \Rightarrow_\Cscr}$ is a sub-graph of $\tup{\Nscr,\Rightarrow \cup \rightarrow}$. $\Cscr$ is a Timed Bundle over $\key{td}$ if:
\begin{enumerate}
  \item $\Cscr$ is finite and acyclic;
  \item $n_2 \in \Nscr_\Cscr$ is Message Input or a Timed Message Input, then there is a unique $n_1 \in \Nscr_\Cscr$ such that $n_1 \rightarrow_\Cscr n_2$;
  \item  $n_2 \in \Nscr_\Cscr$ and $n_1 \Rightarrow n_2$, then $n_1 \in \Nscr_\Cscr$, and $n_1 \Rightarrow_\Cscr n_2$;
  \item $n = \tup{\p,s,\bar{i}}$ is a node such that $term(n)$ is of the form $\key{myif}~{\b}$ or $\key{myif}~{\b~\#~\tc}$ and $\b$ is evaluated to true, then $n \Rightarrow_\Cscr \tup{\p,s,\bar{i}.1}$ and  $n \nRightarrow_\Cscr \tup{\p,s,\bar{i}.2}$; otherwise  $n \Rightarrow_\Cscr \tup{\p,s,\bar{i}.2}$ and $n \nRightarrow_\Cscr \tup{\p,s,\bar{i}.1}$;
  \item the Timed Constraint Set of $\Cscr$ over $\key{td}$ is satisfiable, \ie, there is a substitution $\sigma$, called model of $\TC(\Cscr,\key{td})$, replacing all time variables in $\TC(\Cscr,\key{td})$ by Real numbers so that all inequalities in $\TC(\Cscr,\key{td})$ are true.
\end{enumerate}
\end{definition}

\comment{clt1:  I think we need to require that the timestamps are fresh
on each `strand' in a bundle.

VN: We do not need and probably do not want to enforce this. The Time Constraint Set will handle all that. For example, in the strand $\n_1@\tv_1 \Rightarrow \n_2@\tv_2 \Rightarrow \n_3@\tv$. The Time Constraint Set will force $\tv_2 = \tv_1$ in all its models.
}

\begin{example} The following is a graphical representation for a Timed Bundle using the Distance Bounding Protocol described in Example~\ref{ex:distance-bounding}:
\begin{center}
  \includegraphics[width=0.55\textwidth]{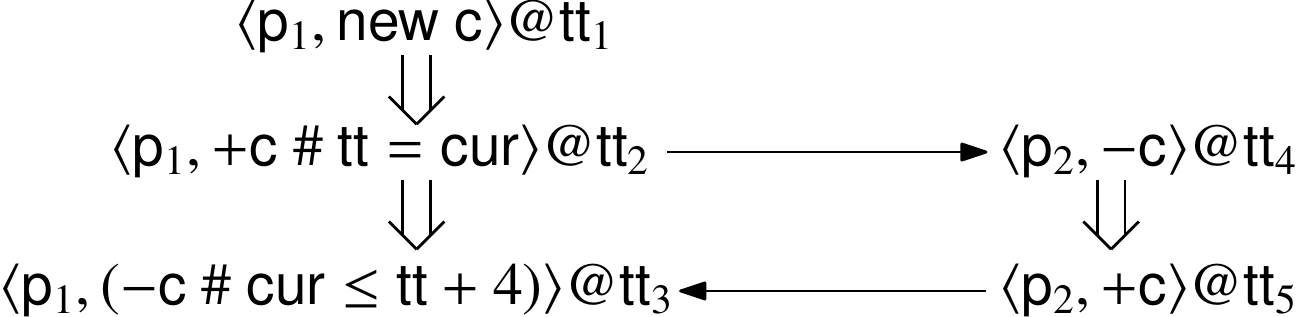}.
\end{center}
It involves two participants $\p_1$ and $\p_2$ which simply exchange a fresh value $\c$.~\footnote{For readability we display graph nodes using the player's id paired with the node term, rather than using the strand identifier and trace position.}
Its Timed Constraint Set should be satisfiable for the assumed Network Topology specified by the function $\key{td}$:
\[
\left\{\begin{array}{c}
\tv_5 \geq \tv_4, \tv_3 \geq \tv_2, \tv_2 \geq \tv_1, \tv = \tv_2, \tv_4 \geq \tv_2 + \td{\p_1}{\p_2},
\tv_3 \geq \tv_5 + \td{\p_2}{\p_1}, \tv_3 \leq \tv + 4 
\end{array}\right\}
\]
Notice that the use of the time symbols in this representation means that
this single object specifies a possibly infinite collection of executions of the Distance Bounding Protocol, where the time symbols are instantiated by concrete timestamps taken from the set of non-negative Real numbers $\mathbb{R}^{+}$. 
This compact representation greatly reduces the state space during automated protocol verification. In our prototype implementation, we use an SMT solver to check whether the set of Time Constraints is satisfiable or not.
\end{example}

\vspace{-2mm}
\section{Timed Intruder Model}
\label{sec:intruder}
\vspace{-2mm}

The Timed Intruder Model is similar to the usual Dolev-Yao Intruder Model in the sense that it can compose, decompose, encrypt and decrypt messages provided it has the right keys. However, unlike the Dolev-Yao intruder, a Timed Intruder is constrained by the physical properties of the systems, namely, an intruder is not able to learn any message instantaneously, instead, must wait until the message arrives.

A Timed Intruder Set is a set of intruder names $\Iscr = \{\ti_1,\ldots,\ti_n\}$ a set of initially known keys $K_P$, which contain all  public keys, all private keys of all the intruders, all symmetric keys initially shared between intruders and honest players, and may contain ``lost keys'' that an intruder learned previously by, for instance, succeeding in some cryptoanalysis. Recall that  Timed Intruders are situated at locations specified by the Network Topology. For instance, $\td{\p_1}{\ti_1} = \key{td}_\Iscr({\p_1},{\ti_1}) = 4$ denotes that the timed needed for a message to travel from  participant $\p_1$ to intruder $\ti_1$ is $4$.

\begin{definition}
  An intruder trace is one of the following, where $\ti$ is a Timed Intruder Name, $\tv,\tv_1,\tv_2,\tv_3$ are time variables, and $\m,\m_1,\ldots, \m_n,\m_1',\ldots,\m_p'$ are messages:
  \begin{itemize}
    \item \textrm{Text Message:} $\tup{\ti,+ \t}@\tv$, where $\t$ is a text constant;
    \item Flushing: $\tup{\ti,-\m}@\tv$;
    \item Forward: $\tup{\ti, -\m, +\m}@(\tv_1,\tv_2)$ denoting the strand $\tup{\ti,-\m}@\tv_1 \Rightarrow \tup{\ti,+\m}@\tv_2$;
    \item Concatenation: \\ $\tup{\ti,-\{\m_1,\ldots,\m_n\}, -\{\m_1',\ldots,\m_p'\}, + \{\m_1,\ldots,\m_n,\m_1',\ldots,\m_p'\}}@(\tv_1,\tv_2,\tv_3)$ denoting the strand

    \begin{small}
    \(
      \begin{array}{l}
     \tup{\ti,-\{\m_1,\ldots,\m_n\}}@\tv_1 \Rightarrow  \tup{\ti,-\{\m_1',\ldots,\m_p'\}}@\tv_2 \Rightarrow \tup{\ti,+ \{\m_1,\ldots,\m_n,\m_1',\ldots,\m_p'\}}@\tv_3
       \end{array}
    \)      
    \end{small}
    \item Decomposing: $\tup{\ti,-\{\m_1,\ldots,\m_n\}, +\{\m_1,\ldots,\m_i\},+\{\m_{i+1},\ldots,\m_n\} }@(\tv_1,\tv_2,\tv_3)$ denoting the strand

    \begin{small}
     \(
      \begin{array}{l}
     \tup{\ti,-\{\m_1,\ldots,\m_i,\m_{i+1},\ldots,\m_n\}}@\tv_1 \Rightarrow 
\tup{\ti,+\{\m_1,\ldots,\m_i\}}@\tv_2 \Rightarrow \tup{\ti,+ \{\m_{i+1},\ldots,\m_n\}}@\tv_3
       \end{array}       
     \)      
    \end{small}
    \item Key: $\tup{\ti,+\k}@\tv$ if $\k \in K_P$;
    \item Encryption: $\tup{\ti,-\k,-\m,+\enc(\m,\k)}@(\tv_1,\tv_2,\tv_3)$ denoting the strand

    \(
    \begin{array}{c}
    \tup{\ti,-\k}@\tv_1 \Rightarrow  \tup{\ti,-\m}@\tv_2 \Rightarrow  \tup{\ti,+\enc(\m,\k)}@\tv_3      
    \end{array}
    \)
    \item Decryption: $\tup{\ti,-\k^{-1},-\enc(\m,\k),+\m}@(\tv_1,\tv_2,\tv_3)$.

    \(
    \begin{array}{c}
    \tup{\ti,-\k^{-1}}@\tv_1 \Rightarrow  \tup{\ti,-\enc(\m,\k)}@\tv_2 \Rightarrow  \tup{\ti,+\m}@\tv_3      
    \end{array}
    \)
  \end{itemize}
\end{definition}

As with the the usual Dolev-Yao intruder model as, e.g., in \cite{thayer99jcs}, the Timed Intruder can send text messages and known keys, receive a message, replay a message, concatenate and decompose messages, and finally encrypt and decrypt messages. There are, however, two differences with respect to the usual Dolev-Yao intruder model as defined in \cite{thayer99jcs}. Each node of the trace is associated with an intruder name $\ti$ and a time variable $\tv$. These are necessary for extracting the Time Constraints of a Strand Graph (as described in Definition~\ref{def:timed-constraint-set}), specifying the physical restrictions of the Timed Intruder.

As the time when timed intruders receive and manipulate messages cannot be measured by the protocol participants, they do not have control over the time variables of timed intruder strands. The following assumption captures this intuition:

\paragraph{Time Variable Disjointness Assumption} For any Bundle $\Bscr$, the set of time variables appearing in protocol participant strands in $\Bscr$ is disjoint from the set of time variables appearing in timed intruder strands in $\Bscr$.

\begin{example} 
\label{ex:mafia-attack}
Let us return to the distance bounding protocol described in Example~\ref{ex:distance-bounding}. The following is an attack, where two colluding intruders $\ti_1$, who is close to $\p_1$, and $\ti_2$, who is close to $\p_2$, collude by sharing a fast channel to fool $\p_1$ into thinking that $\p_2$ is closer than he actually is.
\begin{center}
  \includegraphics[width=0.8\textwidth]{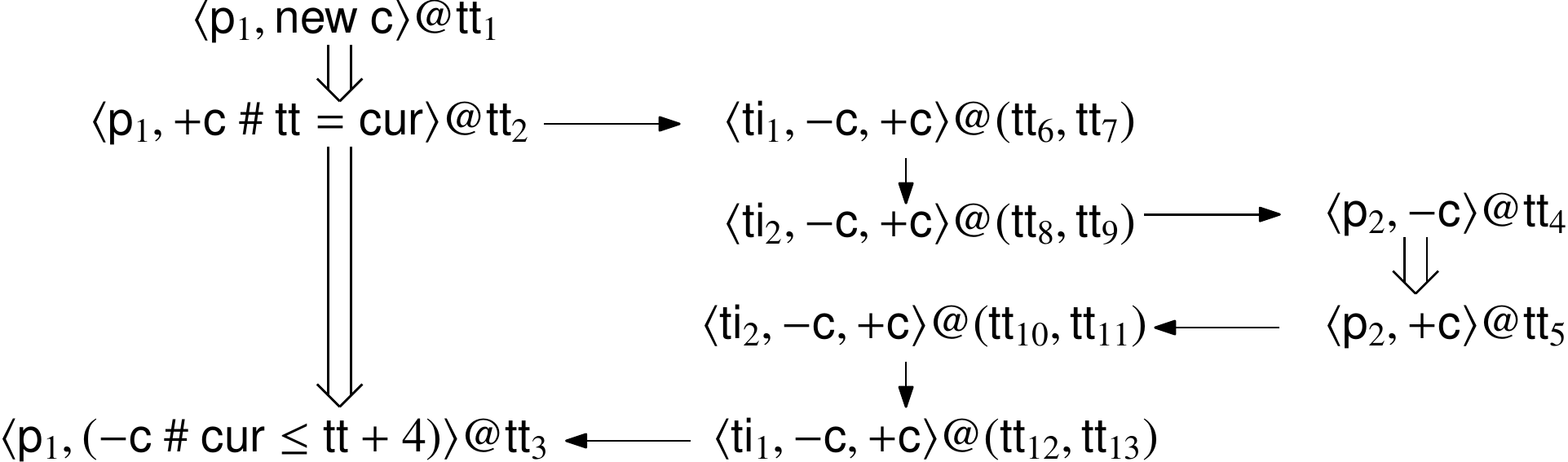}.
\end{center}
The intruders $\ti_1$ and $\ti_2$ simply forward messages between each other and the players $\p_1$ and $\p_2$. However, this is a Bundle only if the following Time Constraint Set is satisfiable:
\[
\left\{
\begin{array}{c}
  \tv_2 \geq \tv_1, \tv = \tv_2, \tv_6 \geq \tv_2 + \td{\p_1}{\ti_1}, \tv_7 \geq \tv_6, \tv_8 \geq \tv_7 + \td{\ti_1}{\ti_2}, \tv_9 \geq \tv_8,\\
  \tv_4 \geq \tv_9 + \td{\ti_1}{\p_2}, \tv_5 \geq \tv_4,
   \tv_{10} \geq \tv_5 + \td{\p_2}{\ti_1}, \tv_{11} \geq \tv_{10}, \tv_{12} \geq \tv_{11} + \td{\ti_2}{\ti_1},\\ \tv_{13} \geq \tv_{12}, \tv_3 \geq \tv_{13} + \td{\ti_1}{\p_1}, \tv_3 \leq \tv_2 + 4
\end{array}
\right\}
\]
This set of constraints represents a set of concrete executions, where the Timed Intruders $\ti_1$ and $\ti_2$ collude. There is a concrete execution only if the set of Time Constraints is satisfiable, which depends on the Network Topology, that is, on the function $\key{td}$. 
\end{example}

\section{Timed Intruder Completeness}
\label{sec:completeness}
Standard Security Protocol Verification is already very challenging. However, automated verification has been very successful in discovering new attacks. A good part of this success is due to the Dolev-Yao intruder model, which greatly simplifies the design of verification tools. Tools can rely on the important result that just a single Dolev-Yao intruder is enough, in the sense that if there is an attack in the presence of multiple (colluding) Dolev-Yao intruders, then there is also an attack in the presence a single Dolev-Yao intruder~\cite{cervesato02isss}.

Unfortunately, for Cyber-Physical Security Protocols, it is not the case that a single Timed Intruder is enough for verification. Consider the attack illustrated in Example~\ref{ex:mafia-attack}. There may be a great number of Timed Intruders, but none of them situated between $\p_1$ and $\p_2$, as illustrated by Figure~\ref{fig:place-intruders}. In such a scenario there might not be an attack as the round time to receive and return a message between such a display of intruders may never be less than the distance bound ($4$). On the other hand, two strategically placed Timed Intruders, as in the second picture in Figure~\ref{fig:place-intruders}, may lead to an attack. 

Clearly there is an unbounded number of choices based on deciding:
\begin{itemize}
  \item How many Timed Intruders are there?
  \item Where are these Timed Intruders located?
\end{itemize}
This is similar to the challenge in usual security protocol verification of determining how many protocol sessions running in parallel should the scenario have, which is undecidable~\cite{Millen99anecessarily}. Fortunately, we are able to prove a completeness result which answers the two questions above. In order to formalize the completeness statement, we introduce some notation.

\begin{definition} Let $\Bscr$ be a Timed Bundle over the Network Topology $\key{td}$ involving the participants $\Pscr = \{\p_1,\ldots,\p_n\}$ and the Timed Intruders $\Iscr = \{\ti_1, \ldots, \ti_n\}$. The graph $\Bscr$ restricted to participants $\Pscr$, written $\Bscr_\Pscr$, is the graph $\tup{\Nscr_\Bscr^\Pscr, (\Rightarrow_\Bscr^\Pscr \cup \rightarrow_\Bscr^\Pscr)}$ specified as follows: 
\begin{itemize}
  \item $\Nscr_\Bscr^\Pscr$ contains only the nodes in $\Bscr$ belonging to a participant in $\Pscr$, \ie, of the form $\tup{\p,s,\bar{i}}$ where $\p \in \Pscr$;
  \item For two nodes $n_1,n_2$ in $\Nscr_\Bscr^\Pscr$, if $n_1 \Rightarrow n_2$ in $\Bscr$, then $n_1 \Rightarrow_\Bscr^\Pscr n_2$;
\item If $n$ is a node in $\Nscr_\Bscr^\Pscr$ whose term is a message receive, 
$-\m$ or $-\m~\#~\tc$,
and $n'$ is a maximal element of the set of predecessors of $n$ in $\Nscr_\Bscr^\Pscr$ under the relation $(\Rightarrow \cup \rightarrow)^*;\rightarrow$
then $n' \rightarrow_\Bscr^\Pscr n$.  We let $\Pscr(n,\Bscr)$ denote this
set of predecessors.
\end{itemize}
\end{definition}
\comment{clt1:  note the new version of the last bullet.  All your examples work with this definition.   Rather than multiple paths between two nodes,
we have multiple predecessors of a receive node, with a unique path
to each predecessor.

VN: I am not sure about the maximal. I thought it would be the minimal. Another way perhaps more direct of defining it is as follows:

Let $\Rightarrow_\Iscr$ be $\Rightarrow$ restricted to Timed Intruder Strands, that is, strands of the form $\tup{\ti_i,s_1}@\tv_1 \Rightarrow \tup{\ti_i,s_2}@t\v_2$. Then, for two nodes $n_1,n_2$ in $\Nscr_\Bscr^\Pscr$, if $n_1 (\rightarrow \cup \Rightarrow_\Iscr)^+ n_2$ in $\Bscr$, then $n_1 \rightarrow_\Bscr^\Pscr n_2$, where the operator $\cdot^+$ is the transitive closure operator.

clt2:  I fixed my definition to force the last arrow to be \rightarrow.
I think it is equivalent to yours.  I've added a little explanatory text
below.
}
Intuitively, a Bundle restricted to the set of participants specifies the events observable by the participants without including the moves corresponding to the timed intruders.  It includes all the edges of the original bundle
connecting two nodes of $\Nscr_\Bscr^\Pscr$.  The ``maximal predecessor''
in $\Nscr_\Bscr^\Pscr$ is the first element of $\Nscr_\Bscr^\Pscr$ encountered
when following edges in the predecessor direction.  It is maximal in the
partial order on nodes induced by the edges of the bundle. 
Thus the terms of nodes in 
$\Pscr(n,\Bscr)$ contain all the terms used by the intruders to derive
the term at node $n$.

The Bundle shown in Example~\ref{ex:mafia-attack} restricted to the participants $\{\p_1,\p_2\}$ is
\vspace{-2mm}
\begin{center}
  \includegraphics[width=0.5\textwidth]{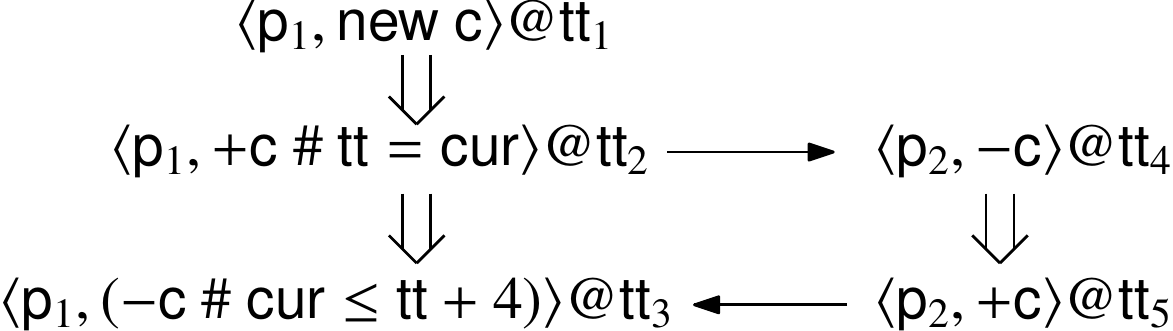}.
\end{center}
The edge $\tup{\p_1,+ \c~\#~\tv = \cur}@\tv_2 \rightarrow \tup{\p_2,- \c}@\tv_4$ in this figure simply specifies that using the message, $\c$, sent by $\p_1$, the timed intruders were able to send the message $\c$ to the participant $\p_2$. 

For another example, consider the following Bundle, where  timed intruder $\ti$ uses his key $\k \in K_P$ and the messages $\c_1$ and $\c_2$ to compose  the message $\enc(\{\c_1,\c_2\},\k)$ to $\p_3$:
\begin{center}
  \includegraphics[width=0.8\textwidth]{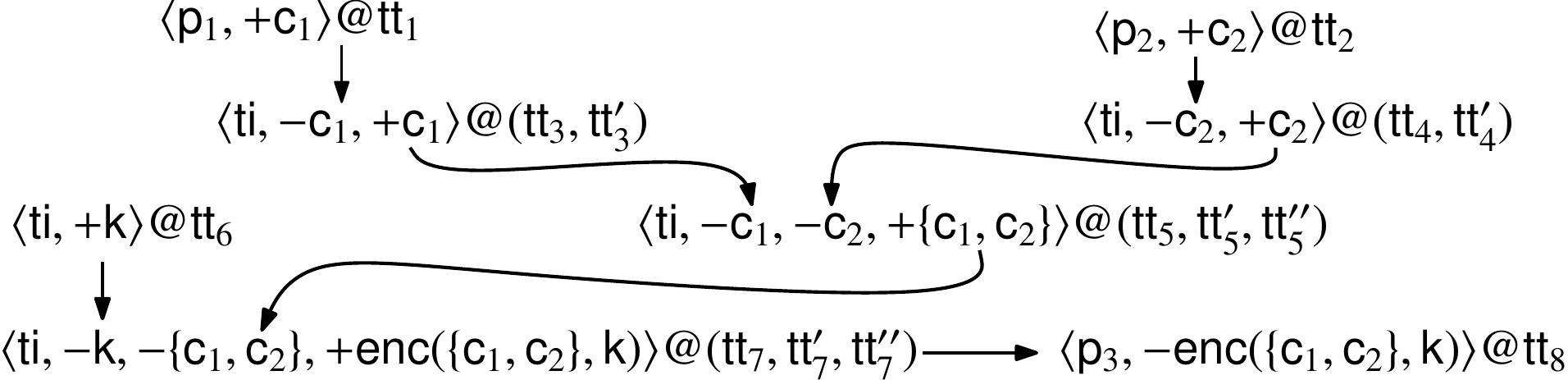}.
\end{center}
The corresponding bundle restricted to the participants $\p_1,\p_2$ and $\p_3$ is:
\begin{center}
  \includegraphics[width=0.8\textwidth]{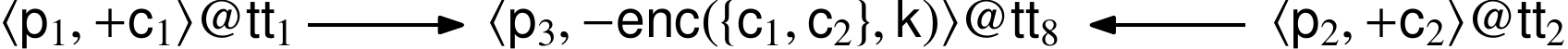}
\end{center}
It captures the fact that the messages sent by $\p_1$ and $\p_2$ are used to generate the message received by $\p_3$ without explicitly showing how intruders manipulated these messages.

\comment{clt: we might want to relate this to the notion of skeletons/shapes

VN: Agreed. From a quick look, they seem different. Shapes and skeletons do not keep track of which messages the intruders used to generate a message, while here we will need this.}

Notice that unlike bundles, a receive node in a restricted bundle may
have multiple incoming edges, reflecting the possibility of processing
by multiple intruders.

\comment{clt1:   The following is not needed.  See comment and notation above.
Your example above is an example of multiple predecessors.

Notice that in a Bundle $\Bscr$ two nodes $\n_1$ and $\n_2$ may be connected by more than one $\rightarrow$ path. This means that an edge in the restricted bundle of $\Bscr$ may be generated by different paths in $\Bscr$. 

Let $e$ be an edge $\n \rightarrow \n'$ in $\rightarrow_\Bscr^\Pscr$. We define $\Pscr(e,\Bscr)$ the set of all paths $\n\rightarrow \n_1 \rightarrow \n_2 \rightarrow \cdots \rightarrow \n_{j} \rightarrow \n'$ in $\Bscr$.

\red{Add example?}
}

\comment{clt1:   I'm running out of steam.   The following needs some
notation tweaks to replace multiple paths between a pair of nodes
by multiple predecessors of a node.  But I think it all works.}
The next two lemmas follow directly from the definition of Bundles and restricted Bundles.
\begin{lemma}
\label{lem:path-shape}
Let $p = \n \leadsto_1 \n_1 \leadsto_2 \n_2 \leadsto_3 \cdots \leadsto_{j-1} \n_{j} \leadsto_{j} \n'$ be a path from $\n$ in $\Pscr(\n',\Bscr)$ to $\n'$, where $\leadsto_i$ is either $\rightarrow$ or $\Rightarrow$ for $1 \leq i \leq j$. 
Then $p$ is necessarily of the form:
\[
  \tup{\p,snd}@\tv \rightarrow \tup{\ti_1,s_1}@\tv_1 \leadsto_2 \tup{\ti_2,s_2}@\tv_2 \leadsto_2 \cdots \leadsto_{j-1} \tup{\ti_j,s_j}@\tv_j \rightarrow \tup{\p',rcv}@\tv'
\]
where $snd$ is a message send $(+ \m)$ or a timed message send $(+\m~ \#~\tc)$, $rcv$ is a message receive $(- \m)$ or a timed message receive $(-\m~ \#~\tc)$, and for $1 \leq i \leq j$, $\tup{\ti_i,s_i}$ are timed intruder strands.
\end{lemma}

\begin{lemma}
\label{lem:constraint-shape}
Let $\Tscr(\Bscr,\key{td})$ be the Time Constraint Set of $\Bscr$ for a given Network Topology $\key{td}$. Let $p$ be a path in $\Bscr$ as described in Lemma~\ref{lem:path-shape} of the form:
\[
  \tup{\p,snd}@\tv \rightarrow \tup{\ti_1,s_1}@\tv_1 \leadsto_1 \tup{\ti_2,s_2}@\tv_2 \leadsto_2 \cdots \leadsto_{j-1} \tup{\ti_j,s_j}@\tv_j \rightarrow \tup{\p',rcv}@\tv'
\]
Then any satisfying model of $\Tscr(\Bscr,\key{td})$ will also satisfy the constraint:
\[
  \tv' \geq \tv + \td{\p}{\ti_1} + \td{\ti_1}{\ti_2} + \cdots + \td{\ti_{j-1}}{\ti_j} + \td{ti_j}{\p'}.
\]
\end{lemma}

The following specifies the equivalence of two Bundles.

\begin{definition}
Let $\Pscr$ be a set of participants and $\Iscr,\Iscr'$ be two possibly equal sets of Timed Intruders. Let $\key{td}_1 = \key{td}_\Pscr \uplus \key{td}_\Iscr$ and $\key{td}_2 = \key{td}_\Pscr \uplus \key{td}_{\Iscr'}$ be Network Topologies. Then we say that a Timed Bundle $\Bscr_1$ over $\key{td_1}$ is equivalent to a Timed Bundle $\Bscr_2$ over $\key{td_2}$, written $\Bscr_1 \cong_{\key{td}_1}^{\key{td}_2} \Bscr_2$, if their Bundles restricted to $\Pscr$ are (syntactically) identical, \emph{i.e.}, $\Bscr_1^\Pscr = \Bscr_2^\Pscr$.\footnote{It is possible to relax this definition so that they are identical modulo time variable names, but this is not needed here.}
\end{definition}

Intuitively, the condition $\Bscr_1^\Pscr = \Bscr_2^\Pscr$  specifies that for the honest participants the two Bundles are equivalent, although they may have different timed intruders in different locations manipulating messages in different ways. Thus, if such a $\Bscr_1$ constitutes an attack, then $\Bscr_2$ also constitutes an attack. 

\comment{clt1:  did I miss something?  It seems like
$\Bscr_1 \vDash_{\key{td}_1}^{\key{td}_2} \Bscr_2$ iff 
$\Bscr_2 \vDash_{\key{td}_2}^{\key{td}_1} \Bscr_1$.
If so, entailment seems like the wrong term.

VN: You are right. Before I had a condition that if $\Tscr(\Bscr_1)$ is satisfiable then $\Tscr(\Bscr_2)$ is also satisfiable. But then the Time Constraint Set of Bundles are always satisfiable.

I change the notation.
}

\paragraph{Timed Intruder Completeness Problem:} 
\begin{quote}
  Let $\Pscr = \{\p_1,\ldots,\p_n\}$ be a set of participants and $\Iscr = \{\ti_1,\ldots,\ti_m\}$ be a set of timed intruders. Let $\key{td}_\Pscr$ be a Network Topology of the participants. Is there a subset $\Iscr' \subseteq \Iscr$ and $\key{td}_{\Iscr'}$ such that for any $\key{td}_\Iscr$ and any Bundle $\Bscr_1$ over $\key{td}_1 = \key{td}_\Pscr \uplus \key{td}_\Iscr$, there is a Bundle $\Bscr_2$ over $\key{td}_2 = \key{td}_\Pscr \uplus \key{td}_{\Iscr'}$ such that $\Bscr_1 \cong_{\key{td}_1}^{\key{td}_2} \Bscr_2$?
\end{quote}

In other words, given a particular scenario with $\Pscr$ participants and a Network Topology for these participants $\key{td}_\Pscr$, is there a Network Topology $\key{td}{_\Iscr'}$ involving a collection of Timed Intruders $\Iscr'$  that can be used to carry out the same observable events for any other Network Topology $\key{td}_\Iscr$ with a possibly larger number of Timed Intruders? 

If such an $\Iscr'$ and $\key{td}_\Iscr'$ exists then an automated verification tool does not have to guess how many timed intruders there are, and where they are located, but simply can use $\Iscr'$ and $\key{td}_\Iscr'$.

\vspace{-2mm}
\subsection{Completeness Proof}
\label{subsec:completeness}
\vspace{-2mm}

We are given a set of participants $\Pscr = \{\p_1,\ldots,\p_n\}$, a set of Timed Intruders $\Iscr = \{\ti_1, \ldots, \ti_m\}$, and a Network Topology $\key{td}_\Pscr$ specifying the time messages take to travel between participants.

\paragraph{A Solution for the Timed Intruder Completeness Problem:} For our solution, we assume that there are as many timed intruders as participants. If this is not the case, we can safely add more dummy timed intruders. We associate with each participant $\p_i$ one Timed Intruder $\ti_{\p_i}$. Thus:
\[
\Iscr'= \{\ti_{\p_1},\ldots, \ti_{\p_n}\}.
\]

Moreover, we assume that the time a message takes to travel between $\p_i$ to $\ti_i$ is 0 (or negligible). Moreover, the time for a message to travel between two Timed Intruders $\ti_{\p_i}$ and $\ti_{\p_j}$ is the same as the time it takes to travel between their corresponding participants $\p_i$ and $\p_j$. Thus:
\[
\begin{array}{l@{\qquad}l}
\key{td}_{\Iscr'}(\p_i,\ti_{\p_i}) = \key{td}_{\Iscr'}(\ti_{\p_i},\p_i) = 0 & \textrm{for all $\p_i\in \Pscr$};\\
\key{td}_{\Iscr'}(\ti_{\p_i},\ti_{\p_j}) = \key{td}_\Pscr(\p_i,\p_j) & \textrm{for all $\p_i,\p_j \in \Pscr$}.
\end{array}  
\]

The Timed Intruders in $\Iscr'$ collude in the following form: whenever a Timed Intruder $\t_{\p_i}$ learns a message $\m$ sent by $\p_i$, it broadcasts this message $\m$ to the remaining Timed Intruders in $\Iscr'\setminus\{\ti_{\p_i}\}$. For example, the Strand for when $\p_1$ sends a message is then as follows:
\vspace{-5mm}
\begin{center}
\[
\includegraphics[width=0.75\textwidth]{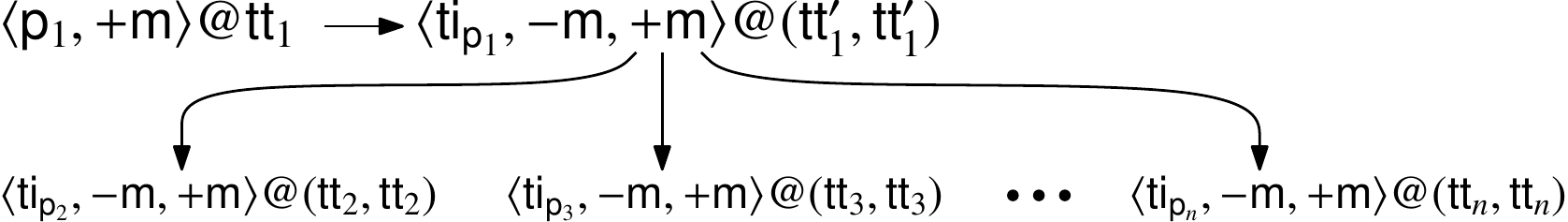}
\]  
\end{center}
Notice that the message $\m$ reaches to a Timed Intruder $\ti_{\p_i}$ at time $\tv_{i}$ which is subject to the Time Constraints $\tv_{i} \geq \tv_{1}' + \key{td}_\Pscr(\p_1,\p_i)$and $\tv_{1}' \geq \tv_1 + \td{\p_1}{\ti_{\p_1}}$, which reduces to $\tv_{1}' \geq \tv_1$ as $\td{\p_1}{\ti_{\p_1}} = 0$. Thus, $\tv_{i} \geq \tv_{1} + \key{td}_\Pscr(\p_1,\p_i)$.
Moreover, if the Timed Intruder $\ti_{\p_i}$ forwards this message to the participant $\p_i$, then this message will be received at a time $\tv_{i}' \geq \tv_{1} + \key{td}_\Pscr(\p_1,\p_i)$, that is, as if the message had traveled directly from $\p_1$ to $\p_i$ without passing through intruders $\ti_{p_1}$ and $\ti_{p_i}$.
\comment{clt1:  Should $\tv_{2}$ above  be $\tv_{1}$? 
VN: Yes corrected.
}

\paragraph{Proof} We will now show that the $\Iscr'$ and $\key{td}_{\Iscr'}$ defined above provide a solution for the Timed Intruder Completeness Problem. For this, assume given a $\key{td}_\Iscr$ and a Bundle $\Bscr_1$ over $\key{td}_1 = \key{td}_\Pscr \uplus \key{td}_\Iscr$.

We will construct a Bundle $\Bscr_2$ over $\key{td}_2 = \key{td}_\Pscr \uplus \key{td}_{\Iscr'}$ such that $\Bscr_1 \cong_{\key{td}_1}^{\key{td}_2} \Bscr_2$. We do so by transforming $\Bscr_1$ into $\Bscr_2$.

Let the following be a sub-graph of $\Bscr_1$ restricted to $\Pscr$:
\begin{center}
\includegraphics[width=0.6\textwidth]{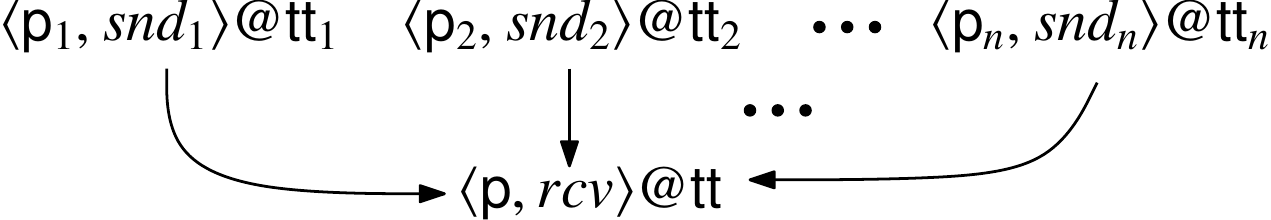}
\end{center}
where for all $1 \leq i \leq n$, $snd_i$ is a Message Output $(+ \m_i)$ or a Timed Message Output $(+\m_i~ \#~\tc)$, $rcv$ is a Message Input $(- \m)$ or a Timed Message Input $(-\m~ \#~\tc)$.

Let $p$ be an arbitrary path from node $\tup{\p_i,snd_i}@\tv_i$ to $\tup{\p,rcv}@\tv$  path in $\Bscr_1$. From Lemma~\ref{lem:path-shape}, $p$ has the shape:
\[
  \tup{\p_i,snd_i}@\tv_i \rightarrow \tup{\ti_1,s_1}@\tv_1 \leadsto_1 \tup{\ti_2,s_2}@\tv_2 \leadsto_2 \cdots \leadsto_{j-1} \tup{\ti_j,s_j}@\tv_j \rightarrow \tup{\p',rcv}@\tv
\]
Moreover, from Lemma~\ref{lem:constraint-shape}, any model satisfying $\Bscr_1$ will also satisfy the constraint:
\begin{equation}
  \tv \geq \tv_i + \td{\p_i}{\ti_1} + \td{\ti_1}{\ti_2} + \cdots + \td{\ti_{j-1}}{\ti_j} + \td{ti_j}{\p}.
  \label{eq:proof1}     
\end{equation}
\comment{clt1:  \tv' ??? where did the prime come from?
Corrected it should be $\tv$.
}
Given our assumption on the Network Topology (Equation~\ref{eq:geometry}), we also have that 
\[
\td{\p_i}{\p} \leq  \td{\p}{\ti_1} + \td{\ti_1}{\ti_2} + \cdots + \td{\ti_{j-1}}{\ti_j} + \td{\ti_j}{\p} 
\]
That is, the time it takes to travel directly from $\p_i$ to $\p$ is less than or equal to the time it takes to travel from $\p_i$ to $\p$ via the timed intruders $\ti_1,\ldots,\ti_j$.

From our solution, we obtain for the sub-graph shown above the following subgraph where all the messages $\m_1,\ldots,\m_n$ are broadcast to all Timed Intruders including the Timed Intruder $\ti_{p}$:
\begin{center}
  \includegraphics[width=0.7\textwidth]{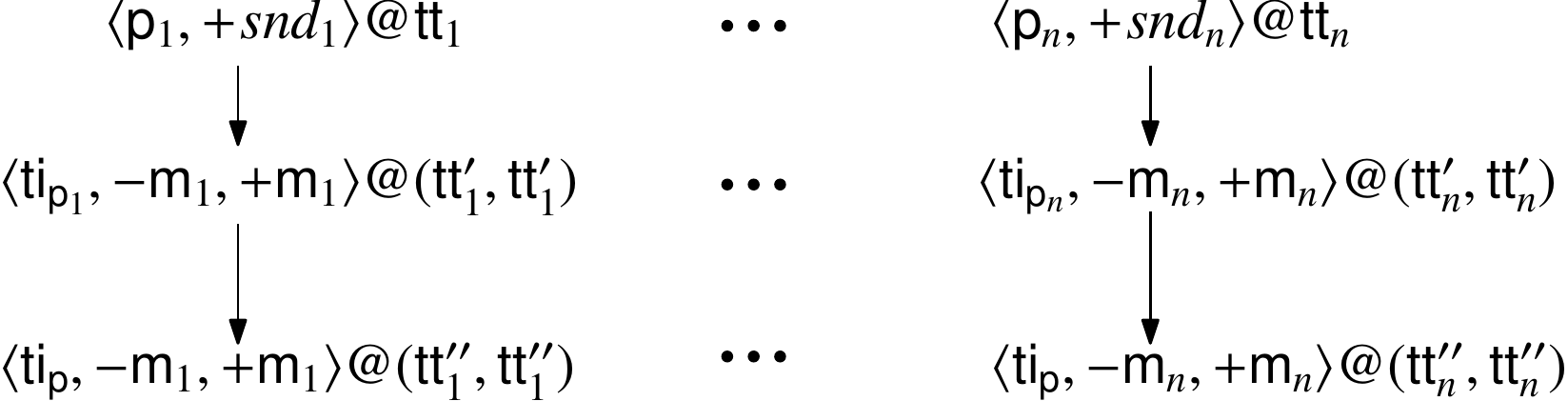}
\end{center}
where the intruder $\ti_{\p}$ receives the messages $\m_1,\ldots,\m_n$. Notice that for $1 \leq i \leq n$, we have that $\tv_i'' \geq \tv_i + \td{\p_i}{\p}$. At this point the intruder $\ti_{\p}$ has all the information he needs to compose the message $\m$. Moreover, he can do so without losing time. Thus he is able to deliver the message $\m$ to $\p$ at time $\tv$ satisfying the constraints:
\begin{equation}
\tv \geq \tv_1 + \td{\p_1}{\p}\quad \tv \geq \tv_2 + \td{\p_2}{\p} \quad \cdots \quad \tv \geq \tv_n + \td{\p_n}{\p}.  
  \label{eq:proof2}
\end{equation}
As any model of the Time Constraints Set of $\Bscr_1$ satisfies Eq.~\ref{eq:proof1}, the same assignment for $\tv_1,\ldots,\tv_n,\tv$ will also satisfy the time constraints in Eq.~\ref{eq:proof2}. Moreover, if any of $snd_1,\ldots,snd_n$ is a Timed Output $(\p_i,+\m_i~\#~\tc_i)$ or $rcv$ is a Timed Input $(\p,- \m~\#~\tc)$ the same assignment will also satisfy $\tc_i$ and $\tc$ because protocol participant strands and timed intruder strands do not share time variable (Time Variable Disjointness Assumption).


By repeating this procedure for each sub-graph in $\Bscr_1$ restricted to $\Pscr$ as shown above, we are able to construct $\Bscr_2$ using $\key{td}_{\Iscr'}$ where the only timed intruder strands are those of the intruders $\Iscr'$ leading to the following result.

\begin{theorem}
  Let $\Pscr$ be participant names and $\Iscr$ be Timed Intruders, such that $|\Iscr| \geq |\Pscr|$. Let $\Iscr'$ and $\key{td}_{\Iscr'}$ be as described above. Then $\Iscr'$ and $\key{td}_{\Iscr'}$ solve the Timed Intruder Completeness Problem.
\end{theorem}

\comment{clt1:   The case when interface nodes have time constraints
is missing. Is that trivial? 
}
\section{Examples and Preliminary Experimental Results}
\label{sec:examples}

We illustrate with some examples that our solution is able to identify attacks on CPSP. We are using the terminology of attacks described in \cite{cremers12oakland}.

\paragraph{External Distance Fraud.} Assume two honest participants $\p_1$ (Verifier) and $\p_2$ (Prover). They exchange some information, normally to authenticate $\p_2$, for example~\cite{santiago16arxiv}, using a standard Needham-Schroeder-Lowe protocol session~\cite{lowe96tacas}, and then carry-out a distance bounding protocol session. The following Timed Strand captures this attack:

\vspace{-2mm}
\begin{center}
  \includegraphics[width=0.75\textwidth]{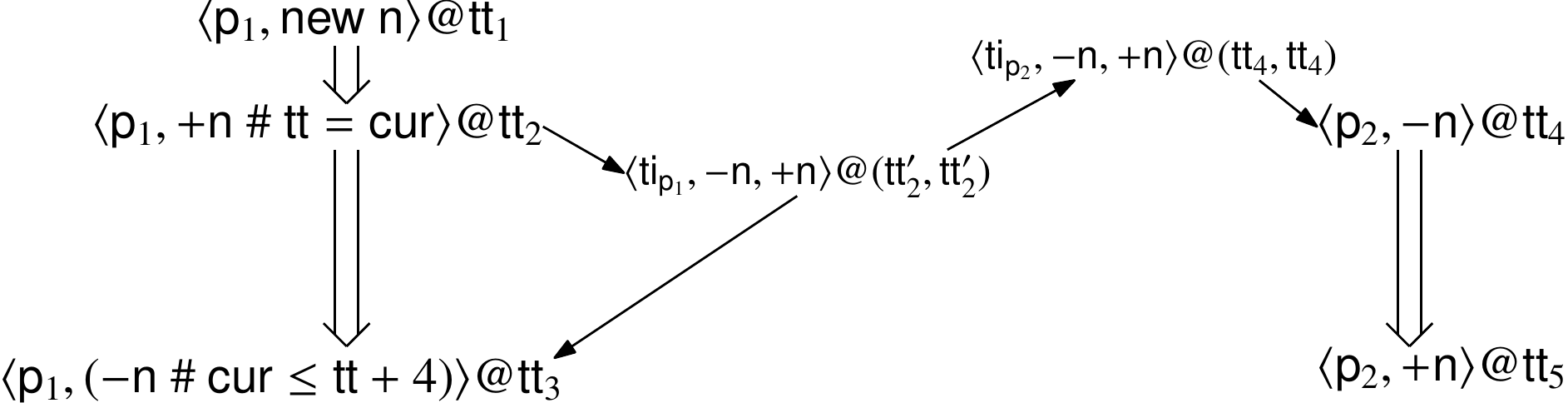}
\end{center}

Notice that the timed intruder $\ti_{\p_1}$ is able to complete the distance bounding session as he is very close to the verifier $\p_1$. This is captured by the Time Constraint Set of this Bundle. Moreover, here we assume that they exchange a nonce, but if we allow equational theories specifying, for example xor operations $\oplus$ as done in \cite{duran-etal-16ijcar}, a similar Timed Bundle would be obtained.

\paragraph{Attack-in-Between-Ticks}
The In-Between-Ticks attack~\cite{kanovich15post} is an instance of a Lone Distance Fraud attack~\cite{cremers12oakland}, where the prover is dishonest but is not colluding with other Timed Intruders. This attack exploits the fact that real verifiers are running on a processor with a slow clock speed. When the verifier receives the response from the prover, he is only able to record the time of receival in the following clock cycle. This is captured by using the Time Constraint $(\floor(\cur) + 1)$ as illustrated by the following Timed Strand:

\vspace{-3mm}
\begin{center}
  \includegraphics[width=0.9\textwidth]{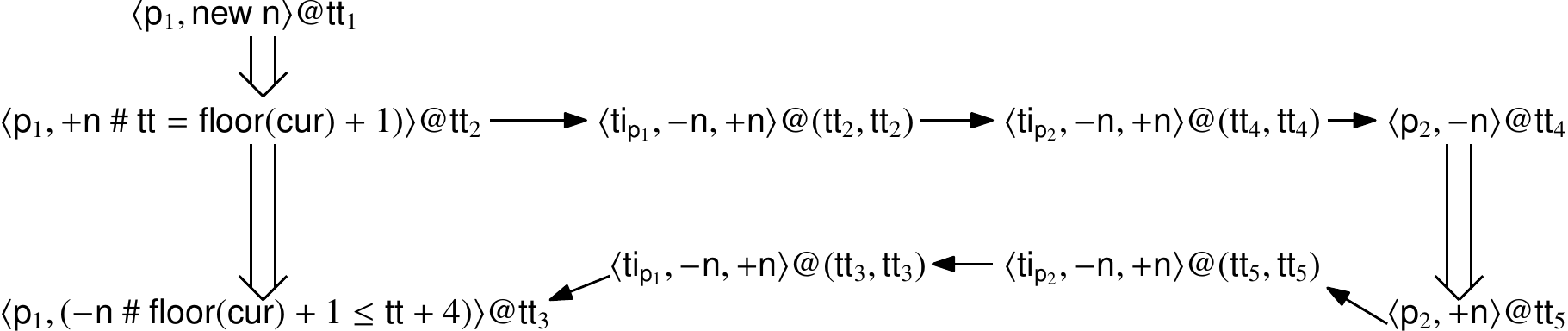}
\end{center}

It is possible to show that the Time Constraint Set of this Timed Strand, $\Tscr$, is satisifiable although the distance between $\p_1$ and $\p_2$ is greater than the distance bound $4$. That is, it is possible to show that the set $\Tscr \cup \{\td{\p_1}{\p_2} > 4, \td{\p_2}{\p_1} > 4\}$ is satisfiable. 



\paragraph{Distance Hijacking} In the Appendix, we show the Timed Bundle with the Distance Hijacking attack described in~\cite{santiago16arxiv} on the protocol that combines the traditional Needham-Schroeder-Lowe protocol and a distance bounding session.  

\vspace{-2mm}
\subsection{Prototype Implementation}
\vspace{-2mm}
We developed a prototype implementation of this strategy in a version of Maude~\cite{clavel-etal-07maudebook} integrated with the SMT solver CVC4~\cite{barrett11cvc4}.
Our preliminary results seem quite promising.

In addition to symbolic time constraints we implemented a symbolic constraint solver in order to tackle the state-space explosion due to the fact that
a timed intruder can generate an unbounded number of messages. It works along the same lines as in usual implementations of such constraint solvers used by tools assuming the standard Dolev-Yao intruder by not instantiating messages generated by the intruder, but rather using symbolic constraints.

\comment{clt2: cite relevant Basin work?
VN: We can do this here or in Related Work.
}

Our prototype used and implements mechanisms for
the main contributions of this paper:
\begin{itemize}
  \item \textbf{Network Topology as a Constraint Set:} While here we assume that the Network Topology is given by a function $\key{td}$ which 
completely determines the time messages take to travel between agents, our implementation allows the user to specify the Network Topology as a set of constraints. For example, the constraint $\td{\p_1}{\p_2} > 4$ specifies the set of Network Topologies where the time it takes for a message to travel from $\p_1$ to $\p_2$ is greater than 4. This reduces even further the decision choices needed when specifying some scenario as one does not need to consider grounded Network Topologies.

  \item \textbf{Time Variables and Time Constraints:} As described here, we use time variables and keep track of the Time Constraints of the constructed Timed Strand, which is initially empty. Whenever a command in our protocol language is executed, we add the corresponding constraint to the set of constraints following Definition~\ref{def:timed-constraint-set}. We then call the SMT solver to check whether the set of constraints is satisfiable. If it is not, then search on this branch of the search tree is aborted. 

  \item \textbf{Timed Intruders:} Our prototype also implements the solution described in Section~\ref{subsec:completeness} for the configuration of timed intruders. This greatly simplifies the number of decisions needed when specifying a verification scenario. Whenever a message is sent by a participant, his corresponding timed intruder broadcasts this message to all other Timed Intruders. A timed intruder is only able to learn such a message when enough time has elapsed. This is implemented also using the SMT solver and adding appropriate time constraints.
\end{itemize}

Table~\ref{tab:exp} summarizes some preliminary experimental results. 

\begin{table}[t]
\begin{center}
\begin{tabular}{@{\quad}c@{\quad}|@{\quad}c@{\quad}|@{\quad}c@{\quad}|@{\quad}c@{\quad}}
\toprule
\textbf{Scenario} & \textbf{Size of Protocols} & \textbf{No of States} & \textbf{Search Time}\\
\midrule
External Distance Fraud & 5 & 12 & 31ms\\
Attack-in-Between-Ticks & 5 & 70 & 55ms\\
Simplified Paywave & 14 & 3224 & 8s\\
Paywave & 22 & 20807 &  78s\\
NSL + Distance Bounding $\star$ & 15 & 86 & 108ms\\
\bottomrule
\end{tabular}
\end{center}
\caption{Preliminary Experimental Results}
\label{tab:exp}
\vspace{-5mm}
\end{table}

The External Distance Fraud and Attack-in-Between-Ticks are as described above. The number of states traversed is quite small for finding these. The distance bounding protocol scheme is used by many other protocols, such as the protocol described in~\cite{santiago16arxiv} (NSL + Distance Bounding) and the lack of its use leads to an attack on the Paywave protocol~\cite{chothia15fcds}. We implemented these to check how our tool scales to larger protocols. We implemented a simplified version of the Paywave protocol omitting some of the steps taken and only concentrating on the core part of the protocol. Our tool was able to find the attack in 8 seconds traversing around 3.2k states. Finally, we implemented the whole Paywave protocol and our tool was also able to find the attack, but now in 78s traversing 20.8k states. 

The use of the SMT solver was essential to reduce the number of states. However, it seems that it is possible to reduce the overhead caused by each call of the SMT solver.

We also experimented with protocols that fall outside  of our language fragment. The NSL + Distance Bounding protocol described in~\cite{santiago16arxiv} with a small modification carries out a standard Needham-Schroeder-Lowe protocol session, followed by a distance bounding protocol using xor. Since our tool does not support yet equational theories, a subject for future work, we modeled the distance bounding session with a pair. Our tool was able to find a terrorist attack in 108 ms traversing 86 states. This attack was not reported in~\cite{santiago16arxiv} as they did not assume that intruders are close to the participants. 

Finally, we also obtained preliminary results on using the tool for checking whether there is a privacy attack on a protocol~\cite{chothia10fcds}. In order to check for such an attack, we need to enumerate all possible executions. (The formal definitions are out of the scope of this paper.) In order to have an idea of how big this set of executions is, we implemented the protocol used for RFID in European passports. The total number of states was only 10 states. This is a promising result for extending this work to check for properties that rely on observational equivalence~\cite{cheval15post}.

\vspace{-2mm}
\section{Related and Future Work}
\label{sec:related}
\vspace{-2mm}

Meadows \etal~\cite{meadows07booktitle} and Pavlovic and Meadows in \cite{pavlovic09spw} 
propose and use a logic called Protocol Derivation Logic (PDL) to formalize and prove the safety of a number of cyber-physical protocols. In particular, they specify the assumptions and protocol executions in the form of axioms, specifying the allowed order of events that can happen, and show that safety properties are implied by the axiomatization used. They do not formalize an intruder model. Another difference between their work and ours is that their PDL specification is not an executable specification.

Another approach similar to \cite{meadows07booktitle}, in the sense that it uses a theorem proving approach, is given by Basin \etal~\cite{basin11iss}. They formalize an intruder model that is similar to ours in Isabelle, and also formalize some cyber-physical security protocols. They then prove the correctness of these protocols under some specific conditions and also identify attacks when some conditions are not satisfied.  Their work  has been a source of inspiration for our intruder model specified in Section~\ref{sec:intruder}. However, they do not propose or investigate the Timed Intruder Completeness Problem.

Chothia \etal~\cite{chothia15fcds} investigate empirically the execution times of commands of CPSP which are carried out by limited resource devices and then, based on these measurements, they propose the inclusion of a distance bounding session to mitigate relay attacks. They proved the security of CPSP by modeling the protocol in different phases. As we illustrate in Example~\ref{ex:passport}, our language allows the inclusion of the measurements themselves. We leave a more detailed analysis to future work.

Cheval and Cortier~\cite{cheval15post} propose a way to prove the observational equivalence with time by reducing it to the observational equivalence based on the length of inputs. They are able to automatically show that RFID protocols used by passports suffer a privacy attack. Their approach is, therefore, different as they do not investigate the Timed Intruder Completeness Problem. Also it is not clear whether from their language one can capture attacks such as the Attack-in-Between Ticks which exploits the time constraints of the verifier. Finally, from our initial experiments with the Passport RFID protocol, we believe that it is also feasible to check for privacy attacks given the very low number of states encountered by our tool. This is left for future work. 

Finally, Malladi \etal~\cite{malladi10corr} formalize distance bounding protocols in strand spaces. 
They then construct an automated tool for protocol verification using a constraint solver to verify a number of examples. There are some similarities 
between their goals and the goal we want to achieve, namely, the automated verification of CPSP and in the use of SMT solvers to do so. However, there are some important differences. Firstly, we formalize and provide a solution to the Timed Intruder Completeness Problem and, secondly, our language seems to have more expressive features, \eg, our time constraints.

The definition of restricted bundle to characterize executions from the protocol participants perspective is inspired by the notions of skeleton and shape in strand space based protocol analysis~\cite{doghmi-guttman-thayer-07tacas,doghmi-guttman-thayer-07mfps}.

Arnaud \etal~\cite{arnaud14ic} propose a model for specifying and reasoning about secured routing protocols where nodes communicate in a direct way with their neighbors. It seems possible to represent our network model using time constraints as they do and not only reason about the routing of packets, but also the time when these arrive, which is important for cyber-physical systems where agents use some routing protocol to communicate. We leave this to future work.

We are currently investigating methods to control even further the state space explosion, for example, using more elaborate symbolic constraint systems for messages and investigating how to support backward Narrowing as in Maude-NPA~\cite{escobar07fosad}. Moreover, we are extending our implementation to support message signatures with equational theories using the library available in Maude~\cite{duran-etal-16ijcar}. Finally, we are investigating definitions of observational equivalence which involve time and that can be implemented using SMT-solvers.


\newpage
\bibliographystyle{abbrv}
\bibliography{master}

\begin{thebibliography}{10}

\bibitem{arnaud14ic}
M.~Arnaud, V.~Cortier, and S.~Delaune.
\newblock Modeling and verifying ad hoc routing protocols.
\newblock {\em Information and Computation}, 238:30 -- 67, 2014.
\newblock Special Issue on Security and Rewriting Techniques.

\bibitem{barrett11cvc4}
C.~Barrett, C.~L. Conway, M.~Deters, L.~Hadarean, D.~Jovanovic, T.~King,
  A.~Reynolds, and C.~Tinelli.
\newblock {CVC4}.
\newblock In {\em Computer Aided Verification - 23rd International Conference,
  {CAV} 2011, Snowbird, UT, USA, July 14-20, 2011. Proceedings}, pages
  171--177, 2011.

\bibitem{basin11iss}
D.~A. Basin, S.~Capkun, P.~Schaller, and B.~Schmidt.
\newblock Formal reasoning about physical properties of security protocols.
\newblock {\em ACM Trans. Inf. Syst. Secur.}, 14(2):16, 2011.

\bibitem{cervesato02isss}
I.~Cervesato.
\newblock Data access specification and the most powerful symbolic attacker in
  {MSR}.
\newblock In {\em Software Security -- Theories and Systems, Mext-NSF-JSPS
  International Symposium, {ISSS} 2002, Tokyo, Japan, November 8-10, 2002,
  Revised Papers}, pages 384--416, 2002.

\bibitem{cheval15post}
V.~Cheval and V.~Cortier.
\newblock Timing attacks in security protocols: Symbolic framework and proof
  techniques.
\newblock In {\em Principles of Security and Trust - 4th International
  Conference, {POST} 2015, Held as Part of the European Joint Conferences on
  Theory and Practice of Software, {ETAPS} 2015, London, UK, April 11-18, 2015,
  Proceedings}, pages 280--299, 2015.

\bibitem{chothia15fcds}
T.~Chothia, F.~D. Garcia, J.~de~Ruiter, J.~van~den Breekel, and M.~Thompson.
\newblock Relay cost bounding for contactless emv payments.
\newblock In {\em Financial Cryptography and Data Security}, 2015.

\bibitem{chothia10fcds}
T.~Chothia and V.~Smirnov.
\newblock A traceability attack against e-passports.
\newblock In {\em Financial Cryptography and Data Security}, pages 20--34,
  2010.

\bibitem{clavel-etal-07maudebook}
M.~Clavel, F.~Dur\'an, S.~Eker, P.~Lincoln, N.~Mart\'i-Oliet, J.~Meseguer, and
  C.~Talcott.
\newblock {\em All About Maude: A High-Performance Logical Framework}, volume
  4350 of {\em LNCS}.
\newblock Springer, 2007.

\bibitem{cremers12oakland}
C.~J.~F. Cremers, K.~B. Rasmussen, B.~Schmidt, and S.~Capkun.
\newblock Distance hijacking attacks on distance bounding protocols.
\newblock In {\em {IEEE} Symposium on Security and Privacy, {SP} 2012, 21-23
  May 2012, San Francisco, California, {USA}}, pages 113--127, 2012.

\bibitem{doghmi-guttman-thayer-07tacas}
S.~F. Doghmi, J.~D. Guttman, and F.~J. Thayer.
\newblock Searching for shapes in cryptographic protocols.
\newblock In {\em Tools and Algorithms for Construction and Analysis of Systems
  ({TACAS})}, volume 4424 of {\em LNCS}, page 523–538. Springer, 2007.

\bibitem{doghmi-guttman-thayer-07mfps}
S.~F. Doghmi, J.~D. Guttman, and F.~J. Thayer.
\newblock Skeletons, homomorphisms, and shapes: Characterizing protocol
  executions.
\newblock In {\em Mathematical Foundations of Program Semantics}, 2007.

\bibitem{duran-etal-16ijcar}
F.~Dur\'an, S.~Eker, S.~Escobar, N.~Mart\'{\i}-Oliet, J.~Meseguer, and
  C.~Talcott.
\newblock Built-in variant generation and unification, and their applications
  in maude 2.7.
\newblock In {\em 8th International Joint Conference on Automated Reasoning},
  2016.

\bibitem{escobar07fosad}
S.~Escobar, C.~A. Meadows, and J.~Meseguer.
\newblock Maude-npa: Cryptographic protocol analysis modulo equational
  properties.
\newblock In {\em Foundations of Security Analysis and Design V, {FOSAD}
  2007/2008/2009 Tutorial Lectures}, pages 1--50, 2007.

\bibitem{kanovich15post}
M.~Kanovich, T.~B. Kirigin, V.~Nigam, A.~Scedrov, and C.~Talcott.
\newblock Discrete vs. dense times in the analysis of cyber-physical security
  protocols.
\newblock In {\em Principles of Security and Trust - 4th International
  Conference, {POST}}, pages 259--279, 2015.

\bibitem{kanovich14fccfcs}
M.~I. Kanovich, T.~B. Kirigin, V.~Nigam, A.~Scedrov, and C.~L. Talcott.
\newblock Towards timed models for cyber-physical security protocols.
\newblock Available in Nigam's homepage, 2014.

\bibitem{lowe96tacas}
G.~Lowe.
\newblock Breaking and fixing the {N}eedham-{S}chroeder public-key protocol
  using {FDR}.
\newblock In {\em TACAS}, pages 147--166, 1996.

\bibitem{malladi10corr}
S.~Malladi, B.~Bruhadeshwar, and K.~Kothapalli.
\newblock Automatic analysis of distance bounding protocols.
\newblock {\em CoRR}, abs/1003.5383, 2010.

\bibitem{meadows07booktitle}
C.~Meadows, R.~Poovendran, D.~Pavlovic, L.~Chang, and P.~F. Syverson.
\newblock Distance bounding protocols: Authentication logic analysis and
  collusion attacks.
\newblock In {\em Secure Localization and Time Synchronization for Wireless
  Sensor and Ad Hoc Networks}, pages 279--298. Springer, 2007.

\bibitem{Millen99anecessarily}
J.~K. Millen.
\newblock A necessarily parallel attack.
\newblock In {\em In Workshop on Formal Methods and Security Protocols}, 1999.

\bibitem{pavlovic09spw}
D.~Pavlovic and C.~Meadows.
\newblock Deriving ephemeral authentication using channel axioms.
\newblock In {\em Security Protocols Workshop}, pages 240--261, 2009.

\bibitem{santiago16arxiv}
S.~Santiago, S.~Escobar, C.~A. Meadows, and J.~Meseguer.
\newblock Effective sequential protocol composition in maude-npa.
\newblock {\em CoRR}, abs/1603.00087, 2016.

\bibitem{thayer99jcs}
F.~J. Thayer, J.~C. Herzog, and J.~D. Guttman.
\newblock Strand spaces: Proving security protocols correct.
\newblock {\em Journal of Computer Security}, 7(1):191--230, 1999.

\end{thebibliography}

\newpage
\appendix

\section{Distance Hijacking Attack}

\begin{center}
\includegraphics[width=0.95\textwidth]{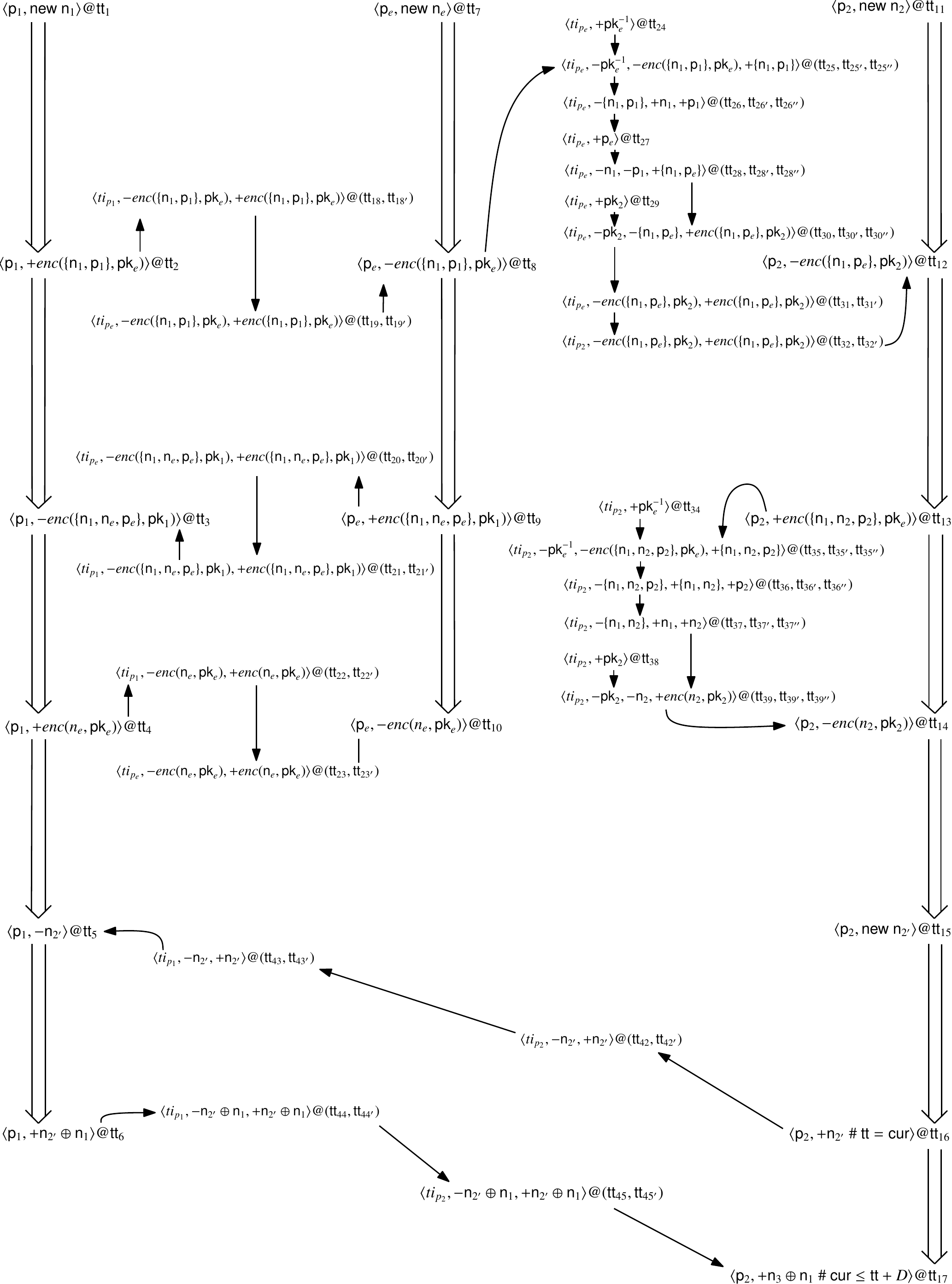}  
\end{center}

\end{document}